\documentclass[12pt]{spie}
\usepackage[utf8]{inputenc}
\usepackage{authblk}
\usepackage{graphicx}
\usepackage{float}
\usepackage{gensymb}
\usepackage{url}
\usepackage{caption}
\usepackage{subcaption}
\usepackage{amssymb}
\usepackage{amsmath}
\usepackage{siunitx}
\usepackage[normalem]{ulem}
\usepackage{wrapfig}
\usepackage{lipsum}
\usepackage{rotating}

\usepackage{pbox}
\usepackage{lineno}

\title{Optimization of telescope focal ratios for MLA-fiber coupled Integral Field Units}

\author[a,*]{Sabyasachi Chattopadhyay}
\author[a,b,c]{Matthew A. Bershady}
\author[b]{Marsha J. Wolf}
\author[b]{Michael P. Smith}

\affil[a]{South African Astronomical Observatory, 1 Observatory Rd, Observatory, Cape Town, 7925, South Africa}

\affil[b]{University of Wisconsin, Department of Astronomy, 475 North Charter Street, Madison, WI 53706, USA}

\affil[c]{Department of Astronomy, University of Cape Town, Private Bag X3, Rondebosch 7701, South Africa}

\authorinfo{Further author information: (Send correspondence to S.C.)\\S.C.: E-mail: sabyasachi@saao.ac.za, Telephone: +27 78 298 6850}

\begin{document} 

\maketitle

\begin{abstract}

We have developed an analytic model for generic image transfer using microlens-coupled fibers to determine the telescope input beam speed that optimizes the lenslet clear aperture and minimizes fiber focal-ratio degradation. Assuming fibers are fed at f/3.5 by the lenslets, our study shows that f/11 is the optimum telescope beam speed to feed a lenslet coupled to a fiber with a 100um diameter core. These considerations are relevant for design of high-efficiency, dedicated survey telescopes that employ lenslet-coupled fiber systems.

\end{abstract}


\section{Introduction}

Large-scale spectroscopic surveys have led to the development of dedicated facilities  using telescopes varying in diameter from a few hundred mm (Konidaris et al. submitted to SPIE proceedings) to 4m (e.g., DESI\cite{desi}, 4MOST \cite{4most}) and even 10m-class (VIRUS\cite{virusp,virusw}, PFS\cite{pfs}, MSE\cite{MSE}). What all of these surveys have in common are multi-object spectroscopic capabilities facilitated by fiber-optic coupling between telescopes and spectrographs. In recent years such surveys have expanded their scope to include multi-object integral field spectroscopy of extended sources to garner a deeper knowledge of nearby galaxies (e.g., SAMI\cite{SAMI} on the 4m AAT and MaNGA\cite{manga} on the 2.5m Sloan Telescope). Again, these surveys use fiber-optic coupling. While there are many formidable, state-of-the-art instruments that use other integral-field techniques (e.g., image-slicing with SPHERE\cite{sphere}, MUSE\cite{vlt} and KMOS\cite{kmos} on VLT, or KCWI\cite{cwi} on Keck), fiber-based integral field units (IFUs; e.g., SparsePak\cite{bershady}, PPak\cite{kelz}, VIRUS-P\cite{virusp}, VIRUS-W\cite{virusw}, MaNGA\cite{drory}, and MEGARA \cite{megara}) are the most flexible and cost-effective, and are our focus here.

Among the various forms of IFUs, microlens-fiber-fed spectrographs have become popular due to the stability, ease of handing, and ease of focal plane formatting for simultaneous, 2D spatial coverage. However, the multi-mode, step-index fibers used always degrade the optical beam in the far-field \cite{arthur,alsmith,carasco}. Unless fibers are fed at fast focal ratios, this phenomenon (focal ratio degradation; FRD) can affect the spectrograph throughput and hence the observational efficiency. On the other hand, the ability of commercial lenslets to produce adequately fast beam speeds while maintaining large clear apertures (relative to their physical aperture) puts constraints on the input beam speed to the lenslets.  We have developed an analytic model for generic image transfer for microlens-coupled fiber IFUs to predict the desirable telescope beam speed given lenslet manufacturing constraints. This model optimizes the trade-offs between lenslet clear aperture and fiber focal-ratio degradation. For this general application, we consider the down-stream spectrograph design either accommodates the fiber output f-ratio, or remodulates the fiber output area--solid-angle product via a second lenslet array at the fiber terminus. Not surprisingly, the limiting factor for the lenslet clear aperture is the lenslet radius of curvature. Consequently, the optimum input focal-ratio to the lenslet depends on the fiber core (and hence lenslet) diameter. Focal-reducers or focal-expanders can always be added to an existing telescope to obtain the optimum input f-ratio. However, for wide-field applications and to achieve the highest system throughput by minimizing optical elements, it is important to consider what native telescope focal-ratio is desirable for microlens array (MLA) coupled fiber IFUs.

\section{The Reimaging integral field unit} 
 
\subsection{Optical Design}

There are two categories of fiber-microlens coupled IFUs: pupil transfer and image transfer. The choice of pupil- vs image-transfer systems reflects where one wants to get the most from scrambling, i.e., in the near-field or far-field. In pupil transfer IFUs, usually a single microlens in an array captures a portion of the focal plane. Plano-convex microlens arrays (PC) were priginally used as the flat surface is easier to position, align, and bond to the fiber. 
The microlens thickness and radius of curvature is adjusted in such a way that the microlens forms its pupil at its back surface where the fiber is bonded. Such a system does not provide an telecentric input beam into the fiber, leading to what is referred to as geometric FRD. There are simple solutions to this problem (e.g., using a bi-convex MLA rather than plano-convex), but in the context of the current analysis we will not consider the general case of pupil-transfer lenslet coupling.

The other option is an image transfer IFU or a reimaging IFU. In this design the simplest approach is two use a pair of microlens arrays back to back. The first microlens produces the pupil at  or near its back surface, just like the pupil-transfer IFU, and effectively acts as a field lens. In this case the first lens is a bi-convex MLA (BC), with the aim here to produce and steer nearly collimated beams from all field angles from the entrance surface subtended by the lenslet element into a second converging lenslet element. This second microlens, with suitable distance, thickness, and radius of curvature (RoC), produces a telecentric image at the exit surface, which can be flat (plano-convex, PC) for optical bonding with the fiber. 

In the image-transfer system, the output micro-image diameter ($d_m$) that is placed on each fiber entrance surface is a free parameter which may be matched to the closest possible step-index, multi-mode fiber diameter. Although it is possible to fill the entire fiber core with the micro-image, our study \cite{chattopadhyay2020} shows that 97-98\% filling is optimum for current options of fiber positioning accuracy which is typically $\pm 3~\mu$m RMS. Our analysis here builds around this optimization.

\begin{figure}[h]
\centering
\includegraphics[width = 0.9\linewidth]{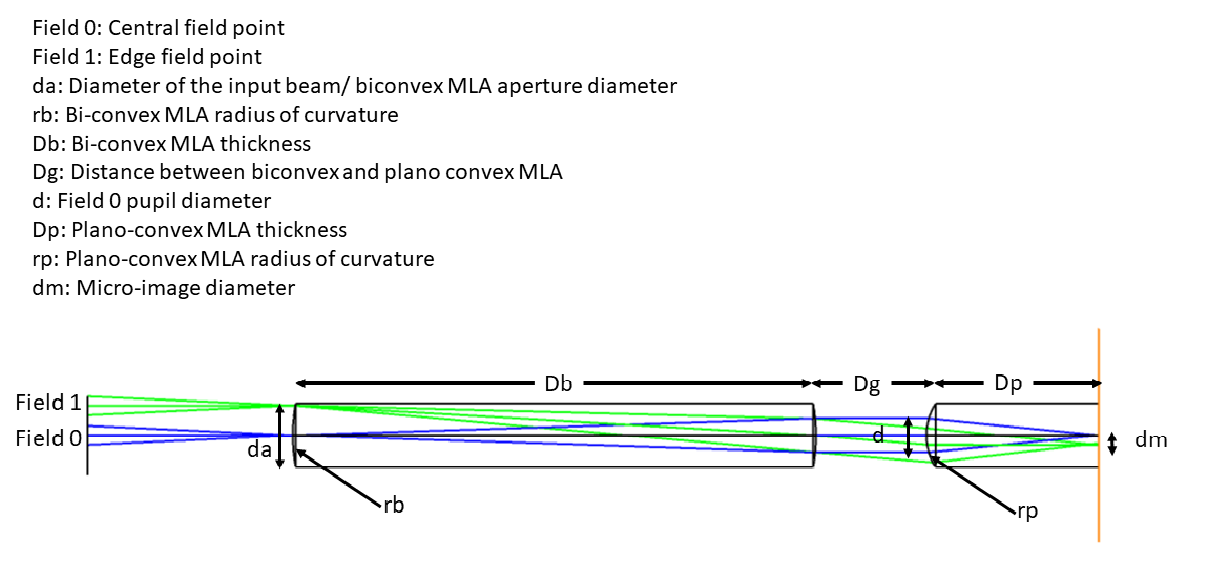}
\caption{Arrangement of a reimaging microlens array system that captures a part of the telescope focal plane ($d_a$) and injects this into the fiber core. Key design parameters are labeled.}
\label{fig:mla}
\end{figure}

In the Figure \ref{fig:mla} we present the building blocks of a reimaging micro-lens system where $f_{\rm tel}$ is the input beam coming from the telescope or a macro-focal reducer/expander. A bi-convex micro-lens of RoC $r_b$, thickness $D_b$, and clear aperture $d_a$ captures this beam. As described earlier, $D_b$ is close to the pupil location formed by the front surface of the first (BC) MLA, and hence if close to half the radius of curvature of that surface. The second lens (plano-convex) transfers this pupil to the image plane, i.e., the PC exit surface. The radius of curvature $r_p$ and thickness $D_p$ are defined to create a specific fiber input f-ratio $f_{\rm fib}$. All the parameters used are listed in table \ref{tab:params}.

\begin{table}[]
\scriptsize
\centering
\begin{tabular}{ccccccccc}
\hline
\begin{tabular}[c]{@{}c@{}}Telescope \\ f-ratio\\ $f_{\rm tel}$\end{tabular} & \begin{tabular}[c]{@{}c@{}}BC MLA \\ RoC\\ $r_b$\end{tabular} & \begin{tabular}[c]{@{}c@{}}BC MLA \\ thickness\\ $D_b$\end{tabular} & \begin{tabular}[c]{@{}c@{}}BC MLA to \\ PC MLA \\ distance\\ $D_g$\end{tabular} & \begin{tabular}[c]{@{}c@{}}PC MLA \\ RoC\\ $r_p$\end{tabular} & \begin{tabular}[c]{@{}c@{}}PC MLA \\ thickness\\ $D_p$\end{tabular} & \begin{tabular}[c]{@{}c@{}}BC MLA \\ clear aperture \\ diameter\\ $d_a$\end{tabular} & \begin{tabular}[c]{@{}c@{}}Central field \\ diameter at \\ PC MLA\\ d\end{tabular} & d/$d_a$ \\ \hline
5  & 0.109 & 0.284  & 0.144 & 0.066  & 0.204 & 0.139 & 0.039 & 0.28 \\
6  & 0.198 & 0.584  & 0.247 & 0.113  & 0.35  & 0.167 & 0.067 & 0.4  \\
7                                                                        & 0.315                                                         & 0.964                                                               & 0.35                                                                            & 0.161                                                         & 0.496                                                               & 0.194                                                                                & 0.094                                                                              & 0.486   \\
8                                                                        & 0.459                                                         & 1.426                                                               & 0.453                                                                           & 0.208                                                         & 0.642                                                               & 0.222                                                                                & 0.122                                                                              & 0.55    \\
9                                                                        & 0.629                                                         & 1.969                                                               & 0.556                                                                           & 0.255                                                         & 0.788                                                               & 0.25                                                                                 & 0.15                                                                               & 0.6     \\
10                                                                       & 0.824                                                         & 2.593                                                               & 0.659                                                                           & 0.302                                                         & 0.934                                                               & 0.278                                                                                & 0.178                                                                              & 0.64    \\
11                                                                       & 1.046                                                         & 3.298                                                               & 0.762                                                                           & 0.35                                                          & 1.079                                                               & 0.306                                                                                & 0.206                                                                              & 0.673   \\
12                                                                       & 1.292                                                         & 4.084                                                               & 0.865                                                                           & 0.397                                                         & 1.225                                                               & 0.333                                                                                & 0.233                                                                              & 0.7     \\
13                                                                       & 1.565                                                         & 4.952                                                               & 0.968                                                                           & 0.444                                                         & 1.371                                                               & 0.361                                                                                & 0.261                                                                              & 0.723   \\
14                                                                       & 1.863                                                         & 5.9                                                                 & 1.071                                                                           & 0.491                                                         & 1.517                                                               & 0.389                                                                                & 0.289                                                                              & 0.743   \\
15                                                                       & 2.186                                                         & 6.929                                                               & 1.174                                                                           & 0.538                                                         & 1.663                                                               & 0.417                                                                                & 0.317                                                                              & 0.76   

\end{tabular}
\caption{Table of parameters used to design the image transfer microlens system for different telescope f-ratio for fiber input beam speed of f/3.5 entering a 100$\mu$m fiber. BC and PC MLA denotes bi-convex and plano-convex microlens array respectively.}
\label{tab:params}
\end{table}

\subsection{Parameter range for microlens designs}

Micro-lens vendors usually define their capabilities in terms of thickness and radius of curvature. In case of the image transfer IFU, both BC and PC RoC are defined by the $f_{\rm fib}$ and $f_{\rm tel}$ as well as the micro-image diameter $d_m$. The $f_{\rm fib}$ also depends on the spectrograph acceptable f-ratio. Modern, wide-field spectrographs typically have input (i.e., collimator) f-ratios in the range of 3 to 5, but they can be faster if the design forgoes significant camera demagnification. For example, Yan et al. (submitted to SPIE manuscript) have shown it is possible to go as fast as f/2.5 and still get significant demagnification even with commercial f/0.9 camera lenses. The slower end of the spectrograph input beam is defined by the acceptable focal ratio degradation (FRD) by the fibers. Studies have shown \cite{Ramsey1988} that slower than a f/5 input beam the FRD becomes significant. We chose $f_{\rm fib}$ = f/3.5 for our design. 

Fiber diameter is usually chosen based on stock choices from 50 to 600 $\mu$m, but custom draws are possible at additional cost. We adopted a standard 100 $\mu$m fiber core for this study. 

The $f_{\rm tel}$ plays a part in determining both the RoC and thickness. Vendors such as {\tt a$\mu$s} are able to fabricate RoC as low as 0.2 mm and a maximum thickness of 6 mm. The RoC limit sets the fastest possible $f_{\rm tel}$ = f/8 while the thickness limitation makes the slowest beam f/14. We have examined the trade-offs over this range to find the optimum telescope beam speed. 

Here we list out all the limits on the parameter space:

\begin{enumerate}
    \item Fiber diameter $d_m$ of 100 $\mu$m.
    \item Fiber input beam speed $f_{\rm fib}$ of f/3.5.
    \item Maximum manufacturing limitation on the thickness ($D_b$ or $D_p$) of a micro-lens of 6 mm.
    \item Fabrication limitation on the minimum radius of curvature of a micro-lens ($r_b$ or $r_p$) of 0.2 mm.
\end{enumerate}

\subsection{Relation between lenslet radius of curvature and clear aperture}

For a simple case of plano-convex lens, there is a relation between the radius of curvature and the clear aperture (CA) of a micro-lens as we discussed in our previous study\cite{chattopadhyay2020}, repeated here:
\begin{equation}
\frac{2-n_g}{n_g-1} \ \geq \ \sqrt{n_g^2 - \left(\frac{d}{2r_p}\right)^2} - \sqrt{1 - \left(\frac{d}{2r_p}\right)^2}.
\label{eq:caroc}
\end{equation}
where $n_g$ is the refractive index of the glass. The relation also holds for a more complex bi-convex microlens. Hence, the manufacturing limit on radius of curvature drives the clear aperture. For an extremely low value of $f_{\rm tel}$, the RoC of both the BC and PC becomes too small to fabricate. Equation \ref{eq:caroc} shows that the clear aperture semi diameter should be about 65\% of the microlens RoC. For all practical purpose the ratio of CA semi-diameter and RoC is kept closer to 50\%. This helps designers to minimize the spherical aberration. Minimizing the spherical aberration also optimizes the field spot sizes (for both the central and edge field) within a tolerable limit\cite{chattopadhyay2020}).

\section{RMS spot size based merit function}

The RMS spot size is a simple metric to judge the goodness of focus and the amplitude of uncorrected aberrations in a system. 
%
For a well focused system (where the central field RMS spot size is minimized), the presence of spherical aberration makes the edge field RMS spot size increase with faster input beam as well as higher radius of curvature of the microlens surface. Thus, the RMS spot size should define the quality of the design. The edge field RMS spot size decreases with increasing telescope f-ratio. This in turn increases the total MLA size, and with
this size increase comes a greater challenge of accurately positioning fibers over the full array. To optimize both the edge and central field RMS spot size together, we used their product as our merit function. Although this is not a unique metric, it does provide a value of $f_{\rm tel}$ close to plausible optimum values which finds close to the minimum RMS spot size for both central and edge field. A different metric could be imagined in a more specific implementation, but this merit function serves as a reasonable general metric for optimization.

\section{Optimization of telescope f-ratio } 
\label{sec:discussion}

\begin{figure}[h]
\centering
\includegraphics[width = 0.9\linewidth]{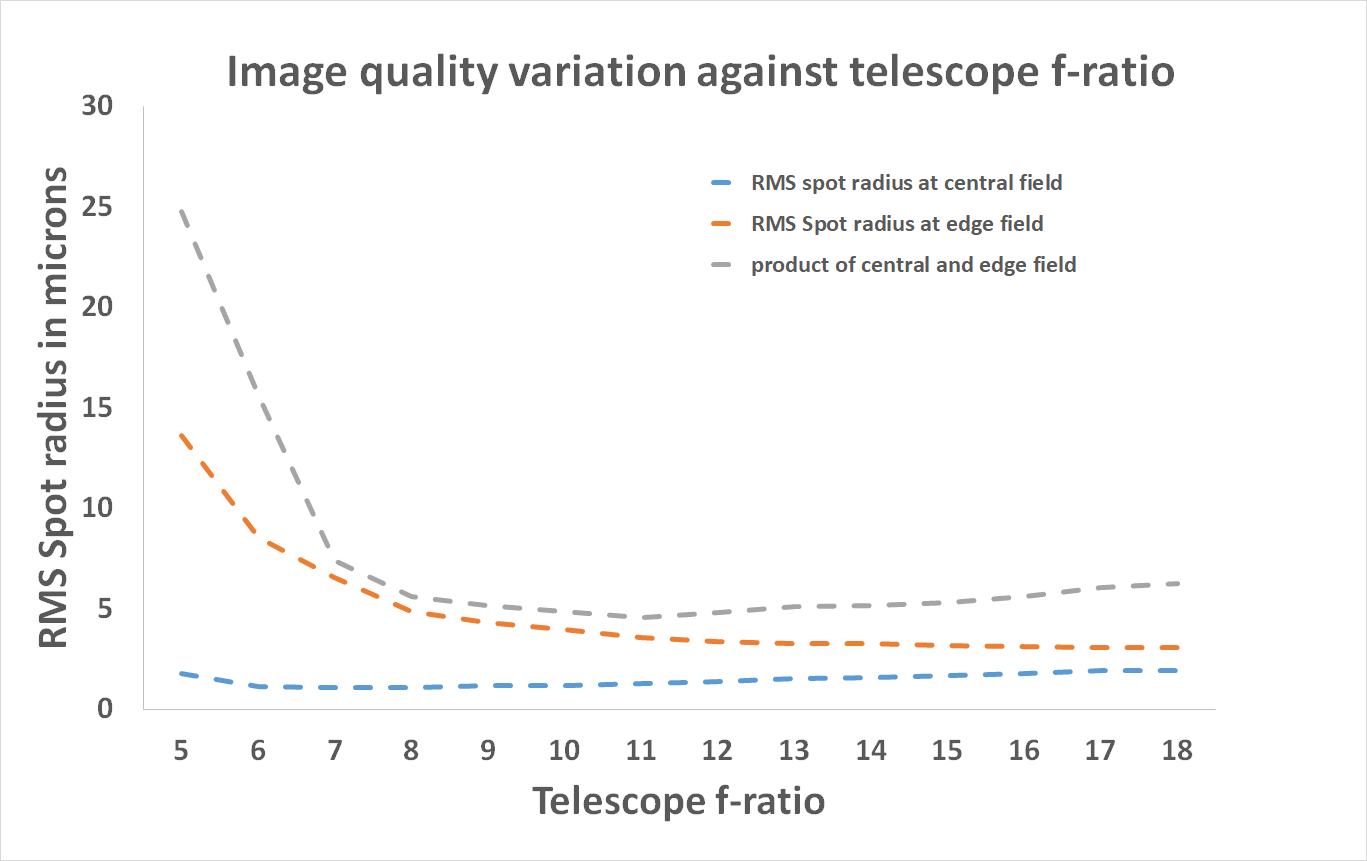}
\caption{Variation in central and edge field RMS spot size for 100$\mu$m fiber with f/3.5 fiber input beam. Our metric for optimum $f_{\rm tel}$ is the product of spot sizes at the central and edge field which is denoted by the grey curve.}
\label{fig:rmsspot}
\end{figure}

Here we considering the image quality at the center and edge of the micro-image field produced from a single lenslet. Figure \ref{fig:rmsspot} shows an example case for a 100 $\mu$m fiber and f/3.5 fiber input beam for a range of $8 < f_{\rm tel} < 14$.  The trend of spot size (RMS radius) at field center, field edge, and their product is given versus telescope f-ratio. The dip at f/11 is the optimization of both field spot sizes. 

\begin{figure}[h]
\centering
\includegraphics[width = 0.9\linewidth]{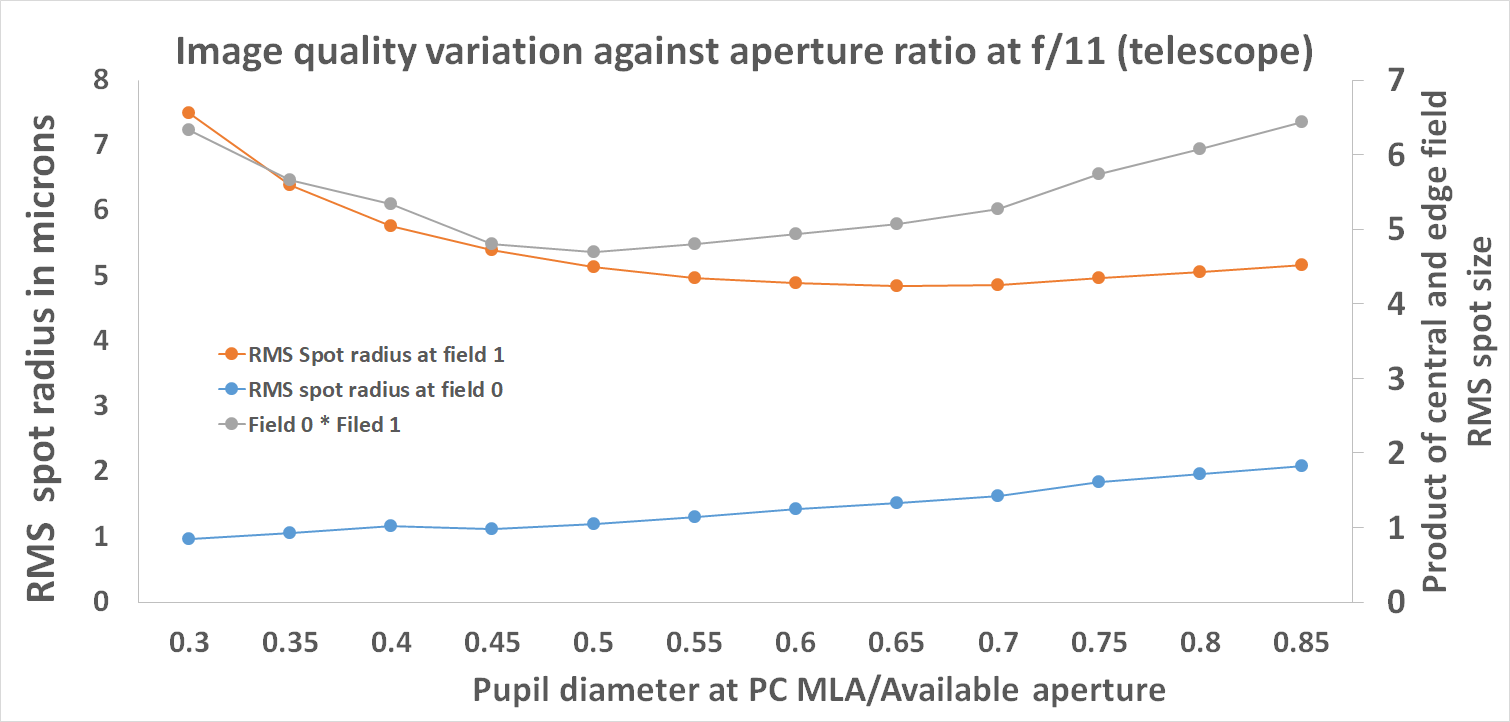}
\caption{Variation in central and edge field RMS spot size against d/$d_a$ for 100$\mu$m fiber with f/3.5 fiber input beam and f/11 telescope beam. Our metric for optimum $f_{\rm tel}$ is the product of spot sizes at the central and edge field which is denoted by the grey curve.}
\label{fig:dda}
\end{figure}

At this point the ``aperture ratio" (d/$d_a$) needs to be optimized to fully define the optical design. Figure \ref{fig:dda} shows the variation of RMS spot size at the micro-image field center, field edge, and the product of the two versust varying aperture ratio at a fixed $f_{\rm tel}=11$. At a given $d_a$, d depends on the BC RoC \& thickness and defines the PC design. Again the field-edge RMS spot size decreases with increasing aperture ratio at the cost of aperture size in turn total MLA size. At d/$d_a = 0.5$ both field spot sizes are optimized. Thus $f_{\rm tel}$ should be roughly 1.5 times of $f_{\rm fib}$.

For the example scenario of 100~$\mu$m fiber fed at f/3.5, it is possible for $f_{\rm tel}$ to be as fast as f/6 and as slow as f/14, but then we are using the maximum available clear aperture (a diameter equivalent to 66\% of the diameter of curvature of the lenslet). Using this much of the clear aperture leads to significant spherical aberration which is indicated by the edge field spot size. As shown in Table \ref{tab:params}, by f/11 we are only using a clear aperture equivalent to 50\% of the diameter of curvature of the lenslet, and this reduces the aberrations to a tolerable level. Going to even slower f-ratios gives marginal improvement in image quality as shown in Figure \ref{fig:rmsspot}. So both d/$d_a$ as well as the product of central and edge field spot sizes point towards the same optimized $f_{\rm tel}$.

Here we try attempt to describe the reason for the optimized telescope f-ratio we find: Slower telescope beams would relax the optical design by increasing the design dimensions (thickness, pitch, RoC) but will create issues in MLA manufacturing and fiber holder dimension: MLA manufacturers (e.g., A$\mu$S) have placed an upper limit on MLA thickness of 6~mm; this puts an upper limit $f_{\rm tel} = f_{\rm high}$ on the telescope f-ratio, e.g., f/14 for $100~\mu$m fiber at $f_{\rm fib}$ = f/3.5. The limiting $f_{\rm tel}$ varies depending on the fiber diameter and $f_{\rm fib}$. The $r_b$ (BC MLA radius of curvature) depends on the ratio d/$d_a$. It is preferable to keep $d$ as high as possible to make the MLA dimensions easier to manufacture. This puts limits on $D_b$. However, beyond $f_{\rm high}$ the required BC MLA thickness cannot be manufactured. The fiber holder fabrication also will be more challenging for increases in pitch with increase in $f_{\rm fib}$. For example at $f_{\rm tel}$ = f/30, the pitch becomes $\sim$1~mm. Absolute positioning accuracy is necessary towards filling optimum fiber core area\cite{sabyasachilitho}. However, fiber pitch of 1 mm would hinder achieving absolute positioning. In such pitch sizes, we would have to rely on relative positioning of fibers. Loosely speaking, the absolute accuracy would be \~dimension $\times$ relative accuracy. Thus, with increasing $f_{\rm tel}$, the achievable absolute positioning accuracy would get worse. Thus the optical design needs to deliver the fastest possible beam speed (to minimize the total lenslet size and to achieve absolute fiber positioning accuracy) but allow for sufficient clear aperture with good image quality.

\section{Summary}

This paper describes a trade study of the input f-ratio fed to compound micro-lens arrays (MLA) that reimage the telescope focal plane onto optical fiber cores. This input f-ratio is either the telescope f-ratio, or a reimaged focus produced by some macro optics. The allowable telescope f-ratios are between f/6 and f/14 for 100 $\mu$m fiber fed at f/3.5, with an optimum value near f/11. The lower bound of f/6 is constrained by the available clear aperture of the MLAs: The required MLA radius of curvature (RoC) decreases with decreasing telescope f-ratio. The upper bound of f/14 is constrained by the maximum thickness of the MLAs (6mm); in general the MLA optics become larger with increasing f-ratio (slower feeds). At f/11 we are only using a clear aperture equivalent to 50\% of the radius of curvature of the lenslet, and this reduces the aberrations to a tolerable level. Going to slower f-ratios gives marginal improvement in image quality. Hence feeding MLAs with beams slower than f/14 should not be exceeded unless critical for some other sub-system. Telescope design choice should target f/11 telescope output beam for spectroscopic instruments for fiber-lenslet coupled IFU.

\acknowledgments 
The University of Wisconsin Madison Graduate School, the U.S. National Science Foundation grant AST 1517006 and AST 1814682, and the South African National Research Foundation SARChI program. 

\bibliography{report} 
\bibliographystyle{spiebib} 

\end{document}